\documentclass[runningheads]{llncs}
\usepackage{graphicx}
\usepackage{tikz}
\usepackage{tkz-graph}
\usepackage{bbm}
\usepackage{booktabs}
\usepackage{listings}
\usepackage{algpseudocode}
\usepackage{algorithm}
\usepackage{amsfonts}
\usepackage{amsmath}
\usepackage{wrapfig}
\usepackage{pgfplots}
\usepackage{pgfplotstable}
\usepackage{upgreek}

\usepackage{graphicx} 
\usepackage{booktabs} 
\usepackage{xparse}   
\NewDocumentCommand{\rot}{O{45} O{1em} m}{\makebox[#2][l]{\rotatebox{#1}{#3}}}%

\usepackage{subfig}

\newtheorem{mydef}{Definition}

\begin{document}

\title{A Novel Token-Based Replay Technique to Speed Up Conformance Checking and Process Enhancement}
\titlerunning{A Novel Token-Based Replay Technique}
%
\author{Alessandro Berti\inst{1,2}\orcidID{0000-0003-1830-4013} \and
Wil van der Aalst\inst{1,2}\orcidID{0000-0002-0955-6940}}
%
%
\institute{Process and Data Science group, Lehrstuhl f\"ur Informatik 9 52074 Aachen, RWTH Aachen University, Germany \and Fraunhofer Gesellschaft, Institute for Applied Information Technology (FIT),\\ Sankt Augustin, Germany}
\maketitle              
\begin{abstract}
Token-based replay used to be the standard way to conduct conformance checking.
With the uptake of more advanced techniques (e.g., alignment based), token-based replay got abandoned. 
However, despite decomposition approaches and heuristics to speed-up computation, the more advanced conformance checking techniques have limited scalability,
especially when traces get longer and process models more complex.
This paper presents an improved token-based replay approach that is much faster and scalable.
Moreover, the approach provides more accurate diagnostics that avoid known problems (e.g., ``token flooding'') and help to pinpoint compliance problems.  
The novel token-based replay technique has been implemented in the PM4Py process mining library.
We will show that the replay technique outperforms state-of-the-art techniques in terms of speed and/or diagnostics.
\end{abstract}

\section{Introduction}

The importance of conformance checking is growing as is illustrated by the new book on conformance checking \cite{conf-check-book-2018} and the Gartner report which states 
``we see a significant trend toward more focus on conformance and enhancement process mining types'' \cite{gartnerpm2018}.
Conformance checking aims to compare an event log and a process model in order to discover deviations and obtain diagnostics information \cite{rogge2016log}.
Deviations are related to process executions not following the process model
(for example, the execution of some activities may be missing, or the activities are not happening in the correct order), and are usually associated to  higher
throughput times and lower quality levels. Hence, it is important to detect them, understand their causes and re-engineer the process in order to avoid
such deviations. A prerequisite for both conformance checking and performance analysis is the ability to replay the event log on the model. This is needed to relate and compare the behavior observed in the log with the behavior observed
in the model. Different replay techniques have been proposed, like {\it token-based replay} \cite{rozinat2008conformance} and {\it alignments} \cite{conf-check-book-2018,adriansyah2011cost}.
In recent years, alignments have become the standard-de-facto technique since they are able to find an optimal match between the process model and a process execution
contained in the event log. Unfortunately, their performance on complex process models and large event logs is poor.
Token-based replay used to be the default technique for conformance checking, but has been almost abandoned in recent years, because the handling of invisible transitions and duplicate transitions require heuristics  to select the proper path in the model.
For example, models may get flooded with tokens in highly non-conforming executions, 
enabling unwanted parts of the process model and hampering the overall
fitness evaluation. Moreover, more detailed diagnostics, that have been developed in recent years, have only been defined in the context of alignments.

In the paper \cite{berti2019reviving}, a revival of token-based replay is proposed.
The approach improves the execution time of the token-based replay operation, increasing the performance gap between token-based replay and alignments (see section \ref{sec:backgroundTokenBasedReplay}). Moreover, 
the approach is able to manage the token-flood problem (see section \ref{sec:tokenFlood}).

This contribution aims to extend \cite{berti2019reviving} in some areas:
\begin{itemize}
\item {\it Root cause analysis} is introduced as a diagnostic (on the output of the token-based replay) provided by the approach.
\item An analysis of a {\it backwards} state-space exploration approach (BTBR) is added. While this technique is not the main contribution of this paper, it provides a viable alternative to the
state-of-the-art approach described in \cite{rozinat2008conformance}. Moreover, some example applications are provided for both BTBR and ITBR.
\item The evaluation section has been extended
and includes now a detailed comparison of fitness values.
\end{itemize}
The remainder of the paper is organized as follows:
in Section~\ref{sec:backgroundSection} an introduction to the main concepts used in this paper is provided.
In Section~\ref{sec:approach}, the problems are defined, and an improved token-based replay is proposed.
In Section~\ref{sec:implementation}, some changes to the implementation are discussed and the evaluation of the approach is proposed.
In Section~\ref{sec:tool}, the tool support is presented, and we elaborate on the additional diagnostics (throughput time and root cause analysis).
In Section~\ref{sec:relatedWorkSection}, the related work is described.
Section~\ref{sec:concl} concludes the paper.

\vspace{-3mm}
\section{Background}
\label{sec:backgroundSection}
\vspace{-1mm}

This section introduces standard concepts related to
Petri nets and event logs. Moreover, the main token-based replay approach \cite{rozinat2008conformance} is introduced.

\vspace{-1mm}
\subsection{Petri Nets}

Petri nets provide a modeling language used from several process mining techniques, e.g., well-known process discovery\footnote{With a process discovery technique, a process model is constructed capturing the behavior seen in an event log. See the book \cite{der2016data} for an introduction to the most popular process discovery algorithms.} algorithms like the alpha miner and the inductive miner \cite{leemans2013discovering} (through conversion of the resulting process tree)
can produce Petri nets. We start from the definition of elementary nets:
\begin{definition}[Nets]
A net is a triple $(P, T, E)$ where:
\begin{itemize}
    \item $P$ and $T$ are disjoint sets of places and transitions respectively.
    \item $E \subseteq (P \times T) \cup (T \times P)$.
\end{itemize}
\end{definition}
Petri nets are such the set of arcs is a multiset over $(P \times T) \cup (T \times P)$.
\begin{definition}[Multiset]
Let $X$ be a set. $B \in \mathcal{B}(X) = X \rightarrow \mathbb{N}$ is a multiset over $X$ where each element $x \in X$ appears $B(x)$ times. Between multisets $B_1 \in \mathcal{B}(X)$ and $B_2 \in \mathcal{B}(X)$ we define the following operations:
\begin{itemize}
\item (union) $B' = B_1 \cup B_2 \iff B'(x) = B_1(x) + B_2(x) ~ \forall x \in X$. We can also say, in the same setting, that $B' = B_1 + B_2$
\item (intersection) $B' = B_1 \cap B_2 \iff B'(x) = \textrm{min}(B_1(x), B_2(x)) ~ \forall x \in X$.
\item (multiset inclusion) $B_1 \leq B_2 \iff B_1(x) \leq B_2(x) ~ \forall x \in X$. Conversely, $B_2 \geq B_1 \iff B_1 \leq B_2$.
\item (difference) $B' = B_1 \setminus B_2 \iff B'(x) = max(B_1(x) - B_2(x), 0) ~ \forall x \in X$. We can also say, in the same setting, that $B' = B_1 - B_2$.
\end{itemize}
Moreover, we say that $x \in B \iff B(x) > 0$.
\end{definition}
An accepting Petri net is a Petri net along with a final marking.
\begin{definition}[Accepting Petri Nets]
A (labeled, marked) accepting {\it Petri net} is a net of the form $PN = (P, T, F, M_0, M_F, l)$ such that:
\begin{itemize}
\item $P$ is the set of {\it places}.
\item $T$ is the set of {\it transitions}.
\item $F \in \mathcal{B}((P \times T) \cup (T \times P))$ is a multiset of arcs.
\item $M_0 \in \mathcal{B}(P)$ is the initial marking\footnote{A marking $M \in \mathcal{B}(P)$  is a place multiset. We denote with $\mathcal{U}_M$ the universe of markings.}.
\item $M_F \in \mathcal{B}(P)$ is the final marking.
\item $l : T \rightarrow {\scriptstyle \sum} \cup \{ \tau \}$ is a {\it labeling function} that assigns to each transition $t \in T$ either a symbol from ${\scriptstyle \sum}$ (the set of labels) or the empty string $\tau$.
\end{itemize}
\end{definition}
\begin{definition}[Preset and Postset of a Place/Transition]
Let $x \in P \cup T$ be a place or a transition. Then
$\bullet x, x \bullet \in \mathcal{B}(P \cup T)$ are defined such that:
\begin{itemize}
    \item $\bullet x(y) = F((y, x)) ~ \forall y \in P \cup T, (y, x) \in (P \times T) \cup (T \times P)$ is the {\it preset} of the element $x$.
    \item $x \bullet(y) = F((x, y))  ~ \forall y \in P \cup T, (x, y) \in (P \times T) \cup (T \times P)$ is the {\it postset} of the element $x$.
\end{itemize}
\end{definition}
The initial marking corresponds to the initial state of a process execution. Process discovery algorithms may associate also a final marking to the Petri net, that is the state in which the process execution should end.
A transition $t$ is said to be {\it visible} if $l(t) \in \sum$; is said to be {\it invisible} if $l(t) = \tau$. If for all $t \in T$ such that $l(t) \neq \tau$, $\arrowvert \{ t' \in T \arrowvert l(t') = l(t) \} \arrowvert = 1$, then the Petri net
contains {\it unique visible} transitions; otherwise, it contains {\it duplicate} transitions.
In the following, some definitions in the context of nets and Petri nets are introduced.
\begin{definition}[Execution Semantics]
The execution semantics of a Petri net is the following:
\begin{itemize}
\item A transition $t \in T$ is {\it enabled} (it may {\it fire}) in $M$ if there are enough tokens in its input places for the consumptions to be possible, i.e. iff $\bullet t \leq M$.
\item Firing a transition $t \in T$ in marking $M$
produces the marking $M' = (M \setminus \bullet t) \cup t \bullet$.
\end{itemize}
\end{definition}
\begin{definition}[Path]
A {\it path} of a net $N = (P, T, E)$ is a non-empty and finite sequence $\eta_1, \ldots, \eta_n$ of nodes of $P \cup T$ such that $(\eta_1, \eta_2), \ldots, (\eta_{n-1}, \eta_n) \in E$.
A path $\eta_1 \ldots \eta_n$ leads from $\eta_1$ to $\eta_n$.
\end{definition}
\begin{definition}[Strongly Connected Nets]
The net $N = (P, T, E)$ is {\it strongly connected}
if a path exists between any node in $P \cup T$, i.e., $(x, y) \in E^{*} ~ \forall ~ x , y \in P \cup T$.
\end{definition}
An important concept is the one of {\it structural components}, that is a collection of {\it subnets} with the property of holding at most one token per time during an execution of the net.
Subnets, S-nets and S-components \cite{esparza1994reduction} are defined as follows.
\begin{definition}[Subnets]
$N' = (P', T', E')$ is a subnet of $N = (P, T, E)$ if $P' \subseteq P$, $T' \subseteq T$ and $E' = E \cap ((P' \times T') \cup (T' \times P'))$.
\end{definition}
\begin{definition}[S-nets]
A net $N' = (P', T', E')$ is an S-net if $\arrowvert \bullet t \arrowvert = 1 = \arrowvert t \bullet \arrowvert$ for every transition $t \in T'$.
\end{definition}
\begin{definition}[S-Component]
A subnet $N' = (P', T', E')$ of $N$ is an S-component of $N$ if $T' = \bullet P' \cup P' \bullet$ and $N'$ is a strongly connected S-net.
\end{definition}

\subsection{Event Logs}

\begin{figure}[ht]
\centering
    \begin{minipage}{0.48\textwidth}
\resizebox{\columnwidth}{!}{%
\begin{tabular}{ll}
\toprule
{\bf Case ID} &                                {\bf Activity} \\
\midrule
       case-10011 &                     Confirmation of receipt \\
       case-10011 &           T02 Check confirmation of receipt \\
       case-10011 &          T03 Adjust confirmation of receipt \\
       case-10011 &           T02 Check confirmation of receipt \\
       case-10017 &                     Confirmation of receipt \\
       case-10017 &      T06 Determine necessity of stop advice \\
       case-10017 &           T02 Check confirmation of receipt \\
       case-10017 &          T03 Adjust confirmation of receipt \\
       case-10017 &           T02 Check confirmation of receipt \\
       case-10017 &  T10 Determine necessity to stop indication \\
       case-10017 &          T03 Adjust confirmation of receipt \\
       case-10017 &           T02 Check confirmation of receipt \\
       case-10017 &          T03 Adjust confirmation of receipt \\
       case-10024 &                     Confirmation of receipt \\
       case-10024 &           T02 Check confirmation of receipt \\
       case-10024 &       T04 Determine confirmation of receipt \\
       case-10024 &  T05 Print and send confirmation of receipt \\
       case-10024 &      T06 Determine necessity of stop advice \\
       case-10024 &  T10 Determine necessity to stop indication \\
       case-10025 &                     Confirmation of receipt \\
       case-10025 &           T02 Check confirmation of receipt \\
       case-10025 &       T04 Determine confirmation of receipt \\
       case-10025 &  T05 Print and send confirmation of receipt \\
       case-10025 &      T06 Determine necessity of stop advice \\
       case-10025 &  T10 Determine necessity to stop indication \\
       case-10028 &                     Confirmation of receipt \\
       case-10028 &           T02 Check confirmation of receipt \\
       case-10028 &       T04 Determine confirmation of receipt \\
       case-10028 &  T05 Print and send confirmation of receipt \\
       case-10028 &      T06 Determine necessity of stop advice \\
       case-10028 &  T10 Determine necessity to stop indication \\
       case-10028 &          T16 Report reasons to hold request \\
       case-10028 &       T17 Check report Y to stop indication \\
       case-10028 &   T19 Determine report Y to stop indication \\
       case-10028 &       T20 Print report Y to stop indication \\
\bottomrule
\end{tabular}
} \\
{\bf a)} Fragment of Event log
    \end{minipage}
	\begin{minipage}{0.01\textwidth}
	~
	\end{minipage}
    \begin{minipage}{0.49\textwidth}
\resizebox{\columnwidth}{!}{%
\includegraphics[width=\textwidth]{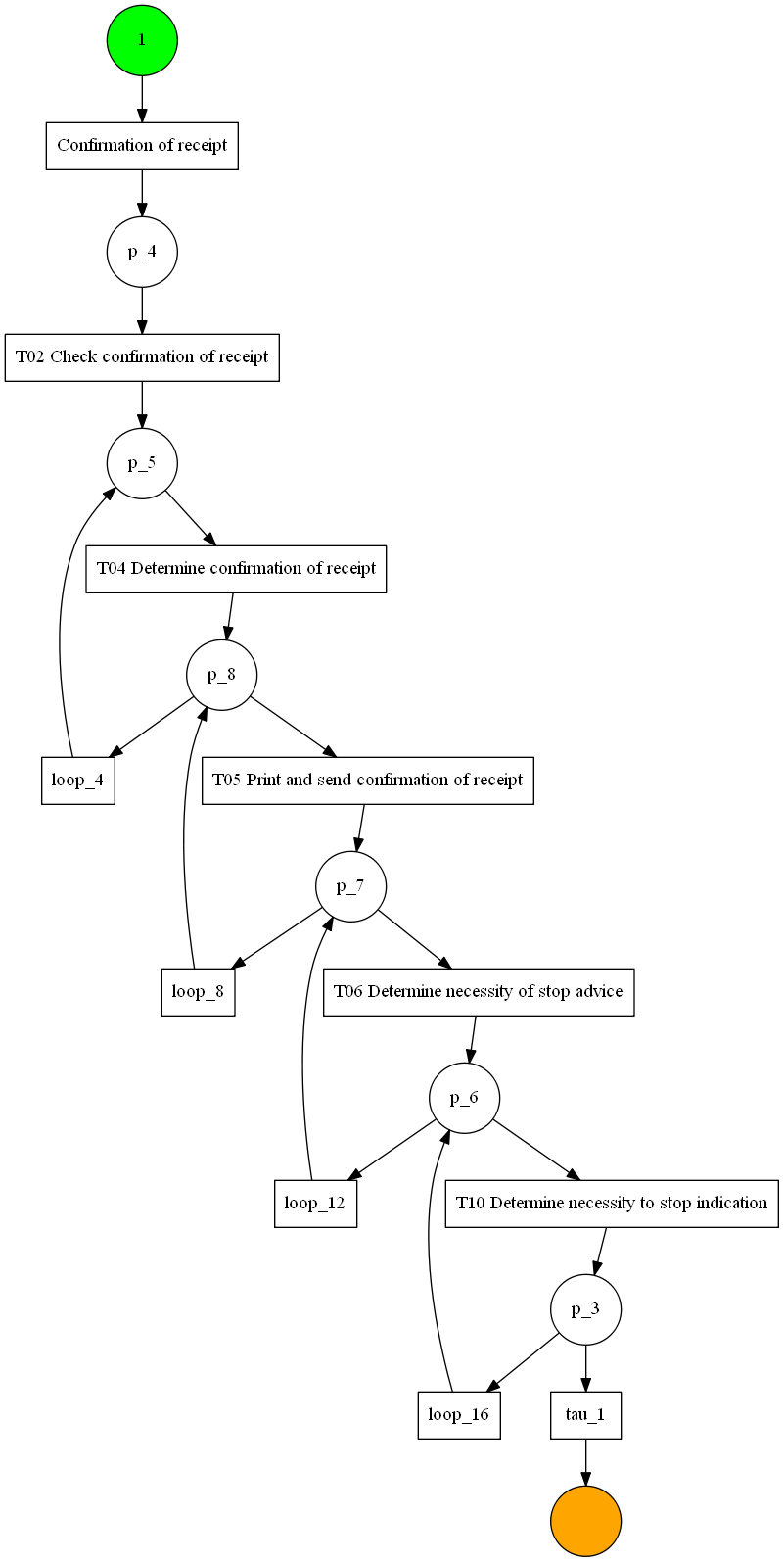}

} \\
{\bf b)} Process model (Petri net)
\end{minipage}
\caption{Petri net extracted by the inductive miner on a filtered version of the ``Receipt phase of an environmental permit application process'' event log.
}
\label{fig:receiptFiltered}
\end{figure}
In process mining, the definition of {\it event log} is fundamental, since it is the input of many techniques as process discovery and conformance checking.
\begin{mydef}[Event Log]
\label{def:classicalEventLogDefinition}
A log is a tuple $L = (C_I, E, \Sigma, \textrm{case\_ev}, \allowbreak \textrm{act}, \allowbreak \textrm{attr}, \leq)$ where:
\begin{itemize}
\item $C_I$ is a set of case identifiers.
\item $E$ is a set of events.
\item $\Sigma$ is the set of activities.
\item $\textrm{case\_ev} \in C_I \rightarrow \mathcal{P}(E) \setminus \{ \emptyset \}$ maps case identifiers onto set of events (belonging to the case).
\item $\textrm{act} \in E \rightarrow \Sigma$ maps events onto activities.
\item $\textrm{attr} \in E \rightarrow (\mathcal{U}_{attr} \not\rightarrow \mathcal{U}_{val})$ (where $\mathcal{U}_{attr}$ is the universe of attribute names, and $\mathcal{U}_{val}$ is the universe of attribute values) maps events onto a partial function assigning values to some attributes.
\item $\leq ~ \subseteq E \times E$ defines a total order on events.
\end{itemize}
\end{mydef}
For a process supported by an information system, an event log is a set of cases, each one corresponding to a different execution of the process. A case contains the list of events
that are executed (in the information system) in order to complete the case. To each case and event, some attributes can be assigned (e.g. the activity and the timestamp at the event level).
A classification of the event is a string describing the event (e.g. the activity is a classification of the event). For each case, given a classification function, the corresponding trace is the list
of classifications associated with the events of the case.
An example application of the inductive miner process discovery algorithm to an event log is represented in Figure \ref{fig:receiptFiltered}.

\subsection{Token-Based Replay}
\label{sec:backgroundTokenBasedReplay}

In process mining, a {\it replay} technique (as introduced in \cite{der2016data}) is a comparison of the behavior of a process execution with the behavior allowed by
a process model. Among the replay techniques, the most important ones are {\it token-based replay} and {\it alignments} that act on Petri nets.
Many different replay techniques are available in the process mining field, targeting different types of process models (not only Petri nets).

Token-based replay is applied to both a trace of the log and an accepting Petri net. The output of the replay
operation is a list of transitions enabled during the replay, along with some numbers ($c$, $p$, $m$ and $r$) defined as follows:
\begin{definition}[Consumed, Produced, Missing, and Remaining Tokens]
Let $L$ be an event log and $\sigma$ be a trace of $L$. Then
$c$ is the number of consumed tokens during the replay of $\sigma$,
$p$ is the number of produced tokens during the replay of $\sigma$,
$m$ is the number of missing tokens during the replay of $\sigma$, and
$r$ is the number of remaining tokens during the replay of $\sigma$.
\end{definition}
At the start of the replay, it is assumed that the tokens in the initial marking are inserted by the environment,
increasing $p$ accordingly (for example, if the initial marking consists of one token in one place, then the replay starts with $p = 1$).
The replay operation considers, in order, the activities of the trace.
In each step, the set of enabled transitions in the current marking is retrieved. If there is a transition corresponding to the current activity, then it is fired,
a number of tokens equal to the sum of the input arcs is added to $c$, and a number of tokens equal to the sum of the output arcs is added to $p$.
If there is not a transition corresponding to the current activity enabled in the current marking, then a transition in the model corresponding to the activity is searched (if there are duplicate corresponding transitions,
then \cite{rozinat2008conformance} provides an algorithm to choose between them).
Since the transition could not fire in the current marking, the marking is modified by inserting the token(s) needed to enable it, and $m$ is increased accordingly.
At the end of the replay, if the final marking is reached, it is assumed that the environment consumes the tokens from the final marking, and $c$ is increased accordingly.
If the marking reached after the replay of the trace is different from the final marking, then missing tokens are inserted and remaining tokens $r$ are set accordingly.
The following relations hold during the replay: $c\leq p + m$ and $m\leq c$. The relation $p + m = c + r$ holds at the end of the replay.
A fitness value can be defined for a trace and for the log.
\begin{definition}[Trace Fitness]
Let $L$ be an event log, $\sigma$ be a trace of $L$, and $c$, $p$, $m$ and $r$ be the consumed, produced, missing and remaining tokens during the replay of $\sigma$. Then the fitness value for the trace is defined as:
$$f_\sigma = \frac{1}{2} \left ( 1 - \frac{m}{c} \right ) + \frac{1}{2} \left ( 1 - \frac{r}{p} \right )$$
\end{definition}
\begin{definition}[Log Fitness]
Let $L$ be a log and $\sigma_0, \ldots, \sigma_n$ be the traces in $L$. Let $c(\sigma_i)$, $p(\sigma_i)$, $m(\sigma_i)$ and $r(\sigma_i)$ be the consumed, produced, missing and remaining tokens during the replay of trace $\sigma_i$.
Then a fitness value for the log $L$ is defined as:
$$f_L = \frac{1}{2} \left (1 - \frac{\sum_{\sigma_i \in L} m(\sigma_i)}{\sum_{\sigma_i \in L} c(\sigma_i)} \right) + \frac{1}{2} \left (1 - \frac{\sum_{\sigma_i \in L} r(\sigma_i)}{\sum_{\sigma_i \in L} p(\sigma_i)} \right )$$
\end{definition}
The log fitness is different from the average of fitness values at trace level.
When, during the replay, a transition corresponding to the activity could not be enabled, and invisible transitions are present in the model, a technique is deployed to traverse the state space (see \cite{rozinat2008conformance}) and possibly reach a marking in which the given transition is enabled.
A heuristic (see \cite{rozinat2008conformance}) that uses the shortest sequence of invisible transitions that enables a visible task is proposed. This heuristic tries to minimize the possibility that the execution of an invisible transition interferes with the future firing of another activity.
A well-known problem for token-based replay is the {\it token flooding problem} \cite{conf-check-book-2018}. Indeed, when the case differs much from the model, a lot of missing tokens are inserted during the replay.
As a result of all the added tokens, many transitions become enabled. Therefore, also deviating events are likely to match an enabled transition.
This leads to misleading diagnostics because unwanted parts of the model may be activated, and so the fitness value for highly problematic
executions may be too high.
To illustrate the token-flooding problem consider a process model without concurrency (only loops, sequences, and choices) represented as a Petri net.
At any stage, there should be at most one token in the Petri net. However, each time there is a deviation, a token may be added, and that leads to a state which was never reachable from the initial state.
The original token-based replay implementation \cite{rozinat2008conformance} was only implemented in earlier versions of the ProM framework (ProM4 and ProM5) and proposes localized metrics on places of the Petri net that help to understand which parts
of the model are more problematic.
To improve performance in the original implementation, a preprocessing step is used to group cases having the same trace. 
In this way, multiple cases having the same trace only need to be analyzed once.

\section{Approach}
\label{sec:approach}

In Section \ref{sec:backgroundTokenBasedReplay}, two problems were analyzed, that led to a relative abandonment of the token-based replay technique (older versions of ProM supported this, but ProM 6 does not):
\begin{enumerate}
\item The slowness in the traversal of invisible transitions, for which an expensive state-space exploration is required.
\item The token flooding problem.
\end{enumerate}

To resolve the first problem, the methodology of exploration of the state-space needs to be changed.
We describe in this section two approaches
for token-based replay that address problem (1).
In Section \ref{sec:backwardsTokenReplaySectionNew}, an alternative approach (BTBR) to the one described in \cite{rozinat2008conformance} is provided, in which a backwards state-space
exploration is performed, instead of a forward state-space exploration. This leads to some advantages in managing common constructs in Petri net models, such as skips/loops,
that are invisible transitions. However, a state-space exploration is still required and, with larger models, this is detrimental.
After the backwards token-based replay, in Section \ref{sec:tokenBasedReplay} a novel technique (ITBR) is introduced.
The improved token-based replay shows good performance results in the assessment.

Moreover, some approaches to solve problem (2) are proposed in Section \ref{sec:tokenFlood}, that exploit the properties of the process model in order to determine which tokens in the replay operation are useful and which are ``superfluous''.

\subsection{Backwards Token-Based Replay}
\label{sec:backwardsTokenReplaySectionNew}
\begin{figure}
\vspace{-3mm}
\centering
\includegraphics[width=\textwidth]{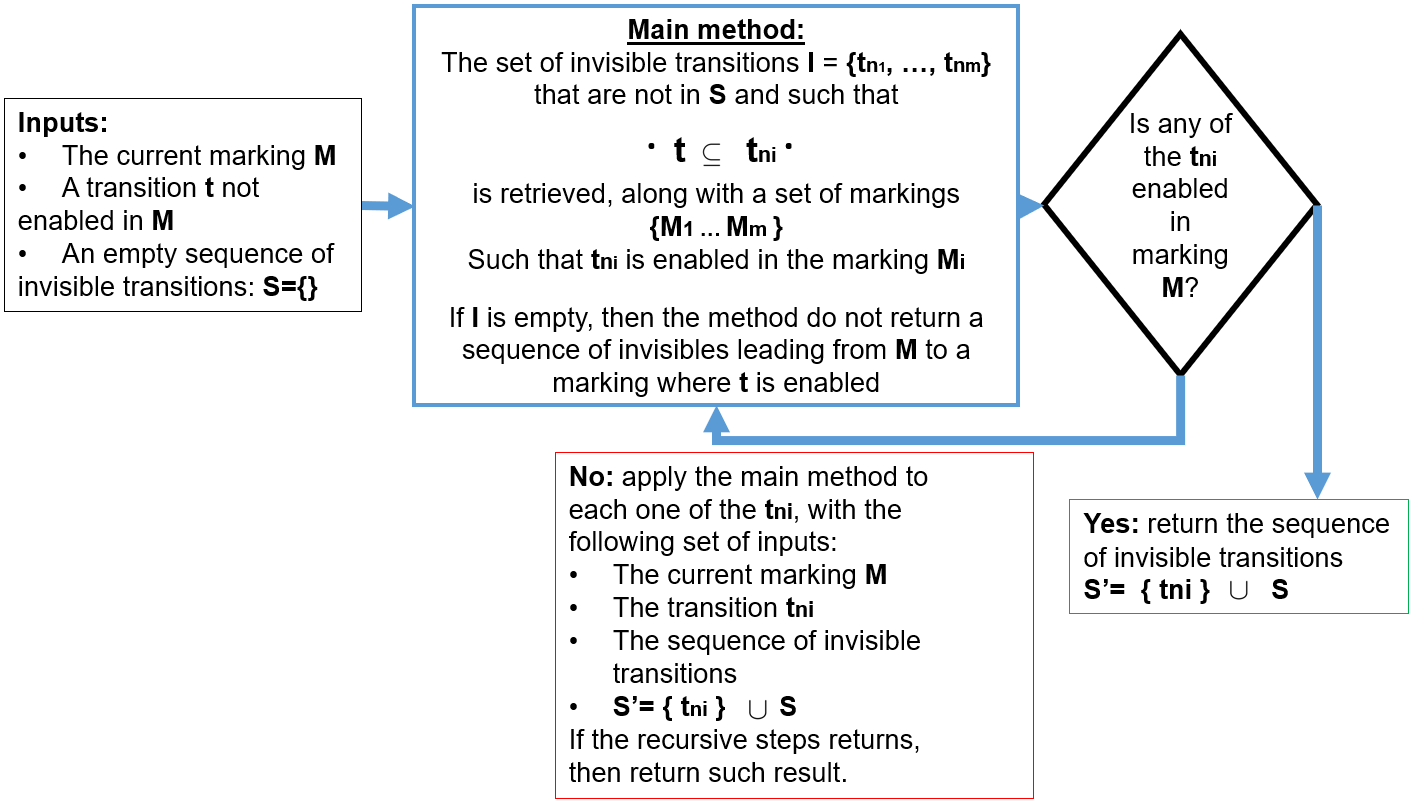}
\caption{A schema of the backwards activation algorithm for invisible transitions.
}
\label{fig:backwardsApproach0}
\vspace{-4mm}
\end{figure}
\begin{figure}
\vspace{-3mm}
\centering
\includegraphics[width=0.85\textwidth]{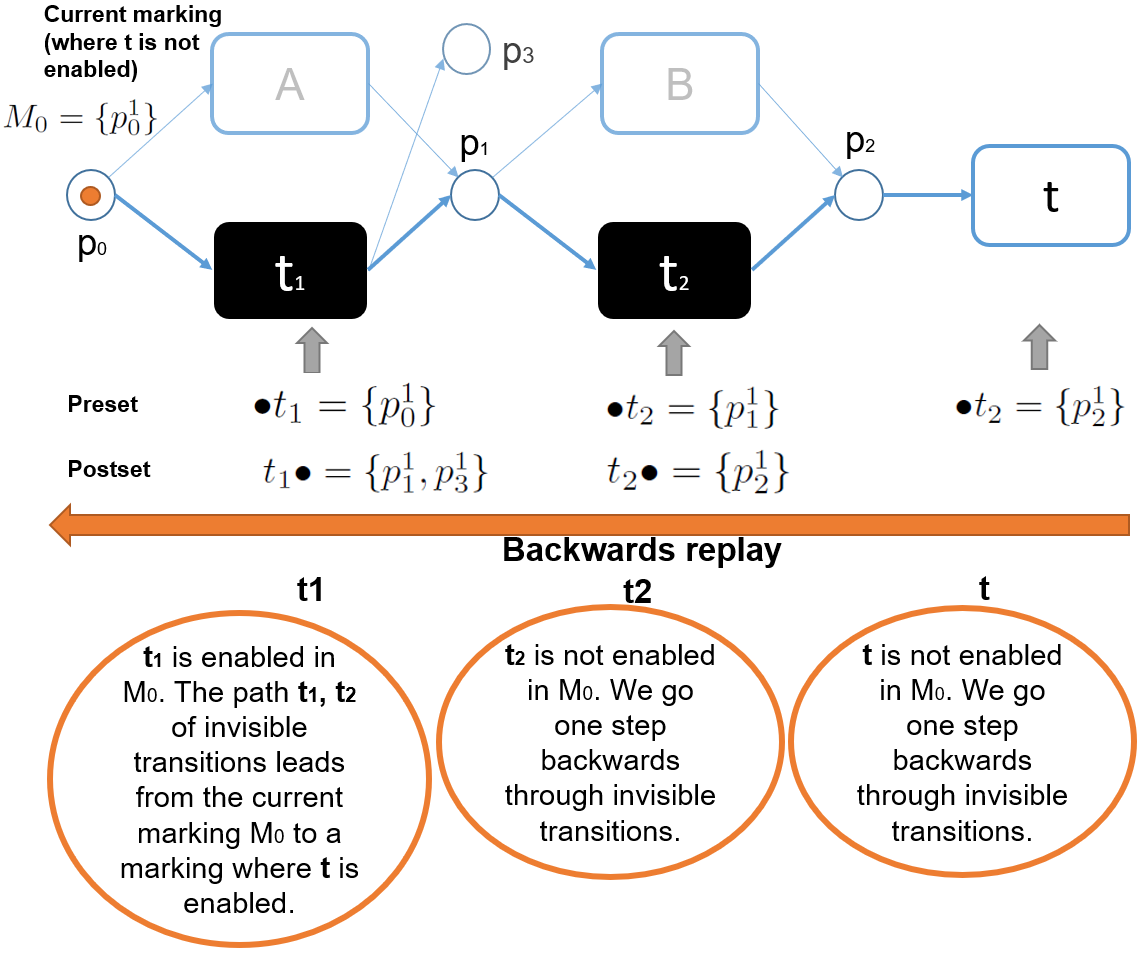}
\caption{An application the backwards activation approach for invisible transitions.
}
\label{fig:backwardsFig}
\vspace{-4mm}
\end{figure}
This section introduces an alternative token-based replay approach that is based on a {\it backwards} state-space exploration. 
The technique adopts the approach presented in \cite{rozinat2008conformance} when a transition corresponding to the replayed activity is enabled in the reached marking.
When no corresponding transitions are enabled in the {\it current marking} $M_0$, an alternative approach (represented in Figure \ref{fig:backwardsApproach0}) is followed to use invisible transitions and reach a marking where at least one corresponding transition is enabled.
In the following, we suppose that $t \in T$ is a transition that corresponds to the replayed activity.
\begin{definition}[Backwards Set of a Transition] \\
Let $PN = (P, T, F, M_0, M_F, l)$ be an accepting Petri net. 
We define the function:
$$B_S : T \rightarrow \mathcal{P}(T)$$
$$B_S(t) = \{ t' \in T ~ \arrowvert ~ l(t') = \tau ~ \wedge ~ \bullet t \leq t' \bullet \}$$
\end{definition}
\begin{definition}[Backwards Marking]
Let $PN = (P, T, F, M_0, M_F, l)$ be an accepting Petri net.
We define the function:
$$M^{\leftarrow} : \mathcal{U}_M \times T \rightarrow \mathcal{U}_M$$
$$M^{\leftarrow}(M, t) = (M \setminus t \bullet) \cup \bullet t$$
As the backwards marking given $M$ and $t$.
The backwards marking is such that $t$ is enabled in $M^{\leftarrow}(M, t)$.
\end{definition}
The idea of the approach is to find a sequence of invisible transitions $t_1, \ldots, t_n$ (where $t_i \neq t_j$ for $i, j \in \{ 1, \ldots, n \}$) such that this sequence leads from the current marking $M_0$ to a marking in which $t$ is enabled.
In order to do so, the state-space is explored going backwards,
and the B-set $B_S(t)$ is considered for further exploration.
In such way, we are sure that the firing of every invisible transition $t_n \in B_S(t)$ leads to a marking where the target transition $t$ is enabled.
If for any $t_n \in B_S(t)$, $M^{\leftarrow}(M, t_n) \subseteq M_0$, then $t_n$ is enabled in marking $M_0$, and leads from $M_0$ to a marking where $t$ is enabled, and the approach stops.
Otherwise, a recursion happens considering each item of the following set of B-sets
$$\{ B_S(t_n) ~ \arrowvert ~ t_n \in B_S(t) \}$$
The recursion continues until a marking that is contained in $M_0$ is reached and a list of transitions leading from $M_0$ to a marking enabling $t$ is obtained.

An example of application of the approach is reported in Figure \ref{fig:backwardsFig}.
There, we need to replay a transition $t$ but we are stuck since we are in a marking $M_0 = \{ p_0^1 \}$ where
$t$ is not enabled. In this case, the approach considers first the invisible transition $t_2$, since $t$ has a preset that is contained in the postset of $t_2$. In doing so,
a marking $M_2$ is found where the invisible transition $t_2$ is enabled. Then, since $M_2$ is not a subset of $M_0$, another backward step is done
and the transition $t_1$ is considered (because its postset contains the preset of $t_2$). A marking $M_1$ is reached where $t_1$ is enabled and, moreover, $M_1 \subseteq M_0$.
This ends the procedure, since from the current marking we are sure to be able to reach a marking where $t$ is enabled by visiting transition $t_1$ and $t_2$: $t_1$ is enabled in $M_0$,
$t_2$ is enabled by construction on the marking obtained firing $t_1$ on $M_0$, and $t$ is enabled by construction on the marking obtained firing $t_2$.

The approach described in this section works nicely with models containing skip/loop transitions. Indeed, while the original token-based replay \cite{rozinat2008conformance} needs to consider
all the possibilities from the current marking, discarding some of them using heuristics, the backwards token-based replay approach considers the minimal marking in which a target transition
is enabled and recursively explores the transitions which postset contains the minimal marking. However, the method is limited in the management of models with concurrency, given the B-set of a transition contains only the invisible
transitions which completely enable that transition.

\subsection{Improved Token-Based Replay}
\label{sec:tokenBasedReplay}

The method described in this part helps to enable a transition $t$ through the traversal of invisible transitions. This helps to avoid the insertion
of missing tokens when an activity needs to be replayed on the model, but no corresponding transition is enabled in the current marking $M$.
Moreover, it helps to avoid time-consuming state-space explorations that are required by \cite{rozinat2008conformance}.
The approach works with accepting Petri nets in which the invisible transitions have non-empty preset and postset;
this because any invisible transition with empty preset/postset would not belong to any shortest path between places.
The description of the method starts from a preprocessing step on the Petri net, and continues with an algorithm to enable transitions using the results of this preprocessing step.

\subsubsection{Preprocessing step}
Given an accepting Petri net $PN = (P, T, F, M_0, M_F, l)$, it is possible to define a directed graph $G = (V,A)$ such that the vertices $V$ are the places $P$ of the Petri net,
and $A \subseteq P \times P$ is such that $(p_1, p_2) \in A$ if and only if at least one invisible transition connects $p_1$ to $p_2$.
Then, to each arc $(p_1, p_2) \in A$, a transition $\uptau(p_1, p_2)$ is associated to, picking one of the invisible transitions connecting $p_1$ to $p_2$.

Using an informed search algorithm for traversing the graph $G$, the shortest paths between nodes are found.
These are sequences of places $\langle p_1, \ldots p_n \rangle$ such that $(p_i, p_{i+1}) \in A$ for any $1 \leq i < n$, and are transformed into sequences
$\langle t_1, t_2, \ldots, t_{n-1} \rangle$ of transitions such that $t_i = \uptau(p_i, p_{i+1})$ for any $1 \leq i < n$.

Given a marking $M$ such that $M(p_1) > 0$ and $M(p_2) = 0$, a marking $M'$ where $M'(p_2) > 0$ could\footnote{During the activation of the sequence, some places could still have missing tokens.} be reached by firing the sequence $\langle t_1, \ldots, t_n \rangle$
that is the shortest path in $G$ between $p_1$ and $p_2$.

\subsubsection{Enabling Transitions}

\begin{figure}
\vspace{-6mm}
\centering
\includegraphics[width=0.9\textwidth]{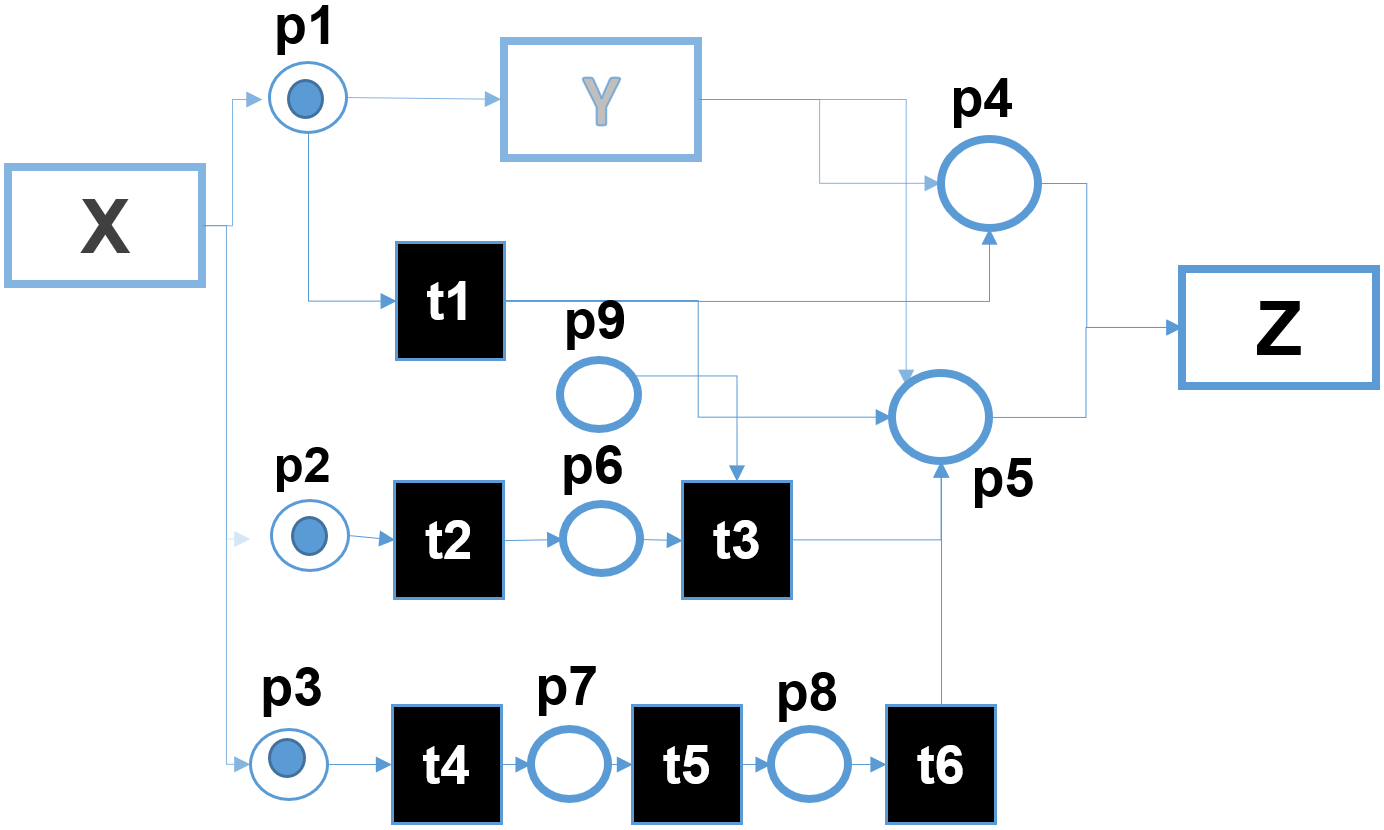}
\caption{A setting in which the application of the improved token-based replay is useful. The replay on this Petri net of the trace $\langle X, Z \rangle$ requires the firing of invisible transitions.}
\label{fig:petrinetCustom}
\end{figure}
This subsection explains how to apply the shortest paths to reach a marking where a transition is enabled. We start from defining the sets $\Delta(M, t)$ and $\Lambda(M, t)$.
\begin{definition}[Delta Set and Lambda Set given a Marking and a Transition]
Let $PN = (P, T, F, M_0, M_F, l)$ be an accepting Petri net. Then we define:
$$\Delta : \mathcal{U}_M \times T \rightarrow \mathcal{P}(P)$$
$$\Delta(M, t) = \{ p \in \bullet t ~ \arrowvert ~ M(p) < F((p,t)) \}$$
and
$$\Lambda : \mathcal{U}_M \times T \rightarrow \mathcal{P}(P)$$
$$\Lambda(M, t) = \{ p \in P ~ \arrowvert ~ F((p, t)) = 0 \wedge M(p) > 0 \}$$
Given a marking $M$ and a transition $t$, $\Delta(M, t)$ is the set of places that miss at least one token to enable transition $t$,
while $\Lambda(M, t)$ is the set of places for which the marking has at least one token and $t$, in order to fire, does not require any of these places .
\end{definition}

Given $\Delta(M, t)$ and $\Lambda(M, t)$,
the idea is about using places in $\Lambda(M, t)$ (that are not useful to enable $t$) and, through the shortest paths, reach a marking $M'$ where $t$ is enabled.

Given a place $p_1 \in \Lambda(M, t)$ and a place $p_2 \in \Delta(M, t)$, if a path exists between $p_1$ and $p_2$ in $G$,
then it is useful to see if the corresponding shortest path
$\langle t_1, \ldots, t_n \rangle$ could fire in marking $M$. If that is the case, a marking $M'$ could be reached, firing such sequence from $M$, that has at least one token in $p_2$.
However, the path may not be a feasible path in the model, or may require a token from one of the input places of $t$.
So, the set $\Delta(M', t)$ may be smaller than $\Delta(M, t)$, since $p_2$ gets at least one token.
The approach is about considering all the combinations of places $(p_1, p_2) \in \Lambda(M, t) \times \Delta(M, t)$ such that a path exists between $p_1$ and $p_2$ in $G$.
These combinations, namely $\{ (p_1, p_2), (p_1', p_2'), (p_1'', p_2''), \ldots \}$, have corresponding shortest paths $S = \{ \langle t_1, \ldots, t_m \rangle, \langle t_1', \ldots, t_n' \rangle, \langle t_1'', \ldots, t_o'' \rangle, \ldots \}$ in $G$.

The algorithm to enable transition $t$ through the traversal of invisible transitions considers the sequences of transitions in $S$, ordered by length, and
tries to fire them. If the path can be executed, a marking $M'$ is reached, and the set $\Delta(M', t)$ may be smaller than $\Delta(M, t)$, since a place in $\Delta(M, t)$ gets at least one token in $M'$.
However, one of the following situations could happen:
\begin{enumerate}
\item no shortest path between combinations of places $(p_1, p_2) \in \Lambda(M, t) \times \Delta(M, t)$ could fire: in that case, we are ``stuck'' in the marking $M$,
and the token-based replay is forced to insert the missing tokens;
\item a marking $M'$ is reached, but $\Delta(M', t)$ is not empty, hence $t$ is still not enabled in marking $M'$. In that case, the approach is iterated on the marking $M'$;
\item a marking $M'$ is reached, and $\Delta(M', t)$ is empty, so $t$ is enabled in marking $M'$.
\end{enumerate}

When situation (2) happens, the approach is iterated. A limit on the number of iterations may be set, and if it is exceeded, then the token-based replay inserts
the missing tokens in marking $M$.
The approach is straightforward when sound workflow nets without concurrency (only loops, sequences, and choices) are considered, since in the considered setting ($M$ marking where transition $t$ is not enabled)
both sets $\Lambda(M, t)$ and $\Delta(M, t)$ have a single element, a single combination $(p_1, p_2) \in \Lambda(M, t) \times \Delta(M, t)$ exists and, if a path
exists between $p_1$ and $p_2$ in $G$, and the shortest path could fire in marking $M$, a marking $M'$ will be reached such that $\Delta(M', t) = \emptyset$ and transition $t$ is enabled.
Moreover, it performs particularly well on models that are output of popular process discovery algorithms (inductive miner \cite{leemans2013discovering}, heuristics miner \cite{weijters2006process}, \ldots) where potentially long chains
of invisible (skip, loop) transitions need to be traversed in order to enable a transition.
The approach described in this subsection can also manage duplicate transitions corresponding to the activity that needs to be replayed.
In that case, we are looking to enable one of the transitions belonging to the set $T_C \subseteq T$ that contains all the transitions corresponding to the activities in the trace.
The approach is then applied on the shortest paths between places.
A similar approach can be applied to reach the final marking when, at the end of the replay of a trace, a marking $M$ is reached that is not corresponding to the
final marking. In that case, $\Delta = \{ p \in P ~ \arrowvert ~ M(p) < M_F(p) \}$ and $\Lambda = \{ p \in P ~ \arrowvert ~ M_F(p) = 0 \wedge M(p) > 0 \}$.
This does not cover the case where the reached marking contains the final marking but has too many tokens.

An example application of the approach is contained in Figure \ref{fig:petrinetCustom}. There, after executing $X$ we have three tokens, one in $p_1$ one in $p_2$ and one in $p_3$. The next replayed activity is $Z$,
that requires one token in $p_4$ and one token in $p_5$. However, since $Y$ is not executed, both tokens are missing. From the marking $\{ p_1, p_2, p_3 \}$
to the set of missing tokens $\{ p_4, p_5 \}$, the set of shortest paths is $S = \{ \langle p_1, p_4 \rangle, \langle p_2, p_6, p_5 \rangle, \langle p_3, p_7, p_8, p_5 \rangle \}$. These are ordered by the length of the path.
Starting from the first path, transition $t_1$ is enabled, and $p_4$ is reached. Then, the second path is examined, however $t_3$ cannot fire hence $p_5$ cannot be reached. So, the last path is executed,
and that leads to putting one token in $p_5$ and eventually enabling $t$.

\subsection{Addressing the Token Flooding Problem}
\label{sec:tokenFlood}
To address the token flooding problem, which is one of the most severe problems when using token-based replay, we propose several strategies.
The final goal of these strategies is to avoid the activation of unwanted transitions that get enabled by the insertion of missing tokens,
keeping the fitness value low for the problematic parts of the model.
The common pattern behind these strategies is to determine \emph{superfluous tokens}, that are tokens that cannot be used anymore.
During the replay, $f$ (initially set to $0$) is an additional variable that stores the number of
``frozen'' tokens. When a token is detected as superfluous, it is ``frozen'': that means, it is removed from the marking and $f$ is increased.
Frozen tokens, like remaining tokens, are tokens that are produced in the replay but never consumed. Hence, at the end of the replay $p + m = c + r + f$.
To each token in the marking, an {\it age} (number of iterations of the replay for which the token has been in the marking without being consumed)
is assigned. The tokens with the highest age are the best candidates for removal.
The techniques to detect superfluous tokens are deployed when a transition requires the insertion of missing tokens to fire, since the marking would then
possibly contain more tokens.
One of the following strategies can be used:
\begin{enumerate}
\item Using a decomposition of the Petri net in semi-positive invariants \cite{martinez1982simple} or S-components \cite{esparza1994reduction,van1996structural}
to restrict the set of allowed markings. Considering S-components, each S-component should hold at most $1$ token, so it is safe to remove the oldest tokens
if they belong to a common S-component.
\item Using place bounds \cite{miyamoto1998calculating}: if a place is bounded to $n$ tokens and during the replay operation the marking contains $m > n$ tokens
for the place, the ``oldest'' tokens according to the age are removed.
\end{enumerate}

\section{Implementation and Evaluation of the Improved Token-Based Replay Technique}
\label{sec:implementation}

In this section, we present some changes to the implementation that have been performed in order to increase the performance of ITBR.
Moreover, we present an assessment of ITBR on real-life logs.

\subsection{Changes to the Implementation to Improve Performance}

In our implementation of token-based replay, we adapt some ideas first used in the context of alignments
\cite{adriansyah2014aligning}:
\begin{enumerate}
\item {\it Post-fix caching}: a post-fix is the final part of a case. During the replay of a case, the couple marking+post-fix is saved in a dictionary along with the list of transitions
enabled from that point to reach the final marking of the model. For the next replayed cases, if one of them reaches exactly a marking + post-fix setting saved in the dictionary, the final
part of the replay is retrieved from the dictionary.
\item {\it Activity caching}: the list of invisible transitions that are activated, from a given marking, to reach a marking where a particular transition is enabled, is saved into a dictionary.
For the next replayed cases, if one of them reaches a marking + target transition setting saved in the dictionary, then the corresponding invisible transitions are fired accordingly to enable the target transition.
\end{enumerate}
In the following:
\begin{itemize}
\item $\textit{CTBR}$ is the classical token-based replay (implemented in ProM 5).
\item $\textit{ITBR}$ is the improved token-based replay described in this paper (implemented in PM4Py).
\item $\textit{ABR}$ is the alignment-based replay (implemented in the ``Replay a Log on Petri Net for Conformance Analysis'' plug-in of ProM 6).
\item $\textit{BTBR}$ is the token-based with backwards state-space exploration described in this paper (implemented in PM4Py).
\item $\textit{AFA}$ is the approach described in \cite{reissner2017scalable} (implemented in PM4Py).
\item $\textit{REABR}$ is the recomposition approach described in \cite{lee2018recomposing} (available in ProM 6).
\item $\textit{ITBR}^{-\textit{PC}}$ is the improved token-based replay without postfix caching
\item $\textit{ITBR}^{-\textit{AC}}$ is the improved token-based replay without activity caching
\item $\textit{ITBR}^{-\textit{PC}/-\textit{AC}}$ is the improved token-based replay without activity or postfix caching.
\end{itemize}

\vspace{-2mm}
\subsection{Evaluation: Execution Time}

\pgfplotstableread[col sep=space,row sep=newline,header=true]{
x   y
34.0 0.28
40.0 0.55
50.0 1.16
102.0 1.80
154.0 2.67
181.0 3.20
244.0 4.30
}\mytable

\pgfplotstableread[col sep=space,row sep=newline,header=true]{
x   y
34.0 0.5
40.0 1.0
50.0 3.50
102.0 6.5
154.0 11.00
181.0 14.90
244.0 28.87
}\mytablee

\begin{table}[ht]
\vspace{-7mm}
\caption{Performance of the different replay approaches on real-life logs and models extracted by the inductive miner. The first columns contain some features of the log. The middle columns
compare ITBR with ABR. In the rightmost columns, the performance of the BTBR, AFA and REABR approaches (BTBR was unable to analyze two of the datasets) are included.}
\centering
\resizebox{\textwidth}{!}{%
\begin{tabular}{|l|cc|cc|c|cc|}
\hline
{\bf Log} & {\bf Cases} & {\bf Variants} & {\bf T.ITBR} & {\bf T.ABR} & {\bf T.BTBR} & {\bf T.AFA} & {\bf T.REABR}\\
\hline
repairEx & 1104 & 77 & 0.06 & 0.2 & 0.04 & {\bf 0.03} & 0.8 \\
reviewing & 100 & 96 & {\bf 0.10} & 0.4 & 0.29 & 0.11 & 2.2 \\
bpic2017(offer) & 42995 & 16 & 0.30 & 1.5 & 0.06 & {\bf 0.01} & 0.18 \\
receipt & 1434 & 116 & {\bf 0.09} & 0.8 & 0.25 & 0.10 & 0.91 \\
roadtraffic & 150370 & 231 & 1.03 & 5.3 & ~ & {\bf 0.09} & 1.37 \\
Billing & 100000 & 1020 & 1.36 & 8.0 & 2.04 & {\bf 1.18} & 9.7 \\
bpic2017(application) & 31509 & 15930 & {\bf 56.1} & 1520.3 & ~ & 116.7 & 1369.2 \\
bpic2018 & 43809 & 28457 & {\bf 145.8} & 8427.2 & 400.99 & 543.01 & 2550.0 \\
bpic2019 & 251734 & 11973 & {\bf 27.0} & 599.1 & 84.60 & 97.50 & 435.9 \\
\hline
\end{tabular}
}
\label{tab:evaluationA}
\vspace{-2mm}
\end{table}

In this section, the improved token-based replay (ITBR) is assessed, looking at the speed and the output of the replay,
against the alignment-based approach on Petri nets (ABR) and the other considered approaches.
Tests contained in Table~\ref{tab:evaluationA} are performed on real-life logs that can be retrieved from the 4TU log repository\footnote{The logs are available at the URL https://data.4tu.nl/repository/collection:event\_logs}.
The tests have been done on an Intel I7-5500U powered computer with 16 GB DDR4 RAM.

\subsubsection{Comparison against ABR}

For real-life logs and models extracted by the inductive miner, the ITBR is $5$ times faster on average.
Even for large logs, the replay time is less than a few seconds.
For the latest BPI Challenge logs, given the model extracted by the inductive miner implementation in PM4Py, there is a noticeable speedup that is $> 20x$,
but also the token-based replay is taking over $20$ minutes.

ABR produces a different output than the one of token-based replay, so results are not directly comparable.
Both are replay techniques, so the goal of both techniques is to provide information about fitness
according to the process model (albeit the fitness measures are defined in a different way, and so are intrinsically different). This is valid in particular
for the comparison of execution times: a trace may be judged fitting according to a process model in a significantly lower amount of time using token-based replay
in comparison to alignments. If an execution is unfit according to the model, it can also be judged unfit in a significantly lower amount of time.
For a comparison, read Section $8.4$ of book \cite{conf-check-book-2018}
or consult \cite{rozinat2005conformance,van2012replaying}.

\subsubsection{Comparison against BTBR}

The comparison between ITBR and BTBR shows that generally ITBR has significantly better performance on larger logs (BPI Challenge 2017 application, BPI Challenge 2018, BPI Challenge 2019).
This shows that the preprocessing step helps to get better performance from token-based replay. The BTBR approach seems also limited in the type of process models it can handle:
while it succeeds for 7 of the considered logs/models, it fails for two settings due to concurrency in the process model.

\subsubsection{Comparison against AFA}

The alignments on finite automaton approach (AFA) shows better performance than ITBR for the vast majority of the logs, excluding the three
bigger logs that have been considered (BPI Challenge 2017 application, BPI Challenge 2018, BPI Challenge 2019).
Moreover, it shows significantly better performance than the other two evaluated alignments approaches (ABR and REABR) working on Petri nets.
Possibly, the worse results of AFA against ITBR in the three BPI Challenge logs have been caused by the bigger size of the automaton, that
can grow fast in complexity.

\subsubsection{Comparison against REABR}

The alignments approach based on a maximal decomposition and, then, a recomposition of the results (REABR) shows a significant performance
increase in comparison to classical alignments (ABR), showing the effectiveness of the approach. However, it records worse results than AFA
that is performed on a different class of models (finite automatons) and ITBR.

\begin{table}[ht]
\vspace{-4mm}
\caption{Comparison of the ITBR execution times on models extracted by the inductive miner on the given logs with or without postfix and activity caching.
Here, the first column is the name of the log, the second is the execution time of ITBR without postfix and activity caching, the third is the execution time of ITBR without activity caching, the fourth
is the execution time of ITBR without the postfix caching, the fifth is the execution time of ITBR with activity and postfix caching enabled.}
\centering
\resizebox{\textwidth}{!}{%
\begin{tabular}{|l|cccc|}
\hline
{\bf Log} & {\bf $\textrm{ITBR}^{-\textrm{PC}/-\textrm{AC}}$(s)} & {\bf $\textrm{ITBR}^{-\textrm{AC}}$(s)} & {\bf $\textrm{ITBR}^{-\textrm{PC}}$(s)} & {\bf $\textrm{ITBR}$(s)} \\
\hline
repairEx & 0.10 & 0.08 & 0.08 & {\it 0.06} \\
reviewing & 0.33 & 0.42 & 0.14 & {\it 0.10} \\
bpic2017(offer) & 0.37 & 0.42 & 0.30 & {\it 0.30} \\
receipt & 0.17 & 0.15 & 0.12 & {\it 0.09} \\
roadtraffic & 1.58 & 2.08 & 1.18 & {\it 1.03} \\
Billing & 2.23 & 1.91 & 1.45 & {\it 1.36} \\
bpic2017(application) & 75.7 & 69.1 & 64.3 & {\it 56.1} \\
bpic2018 & 164.8 & 161.8 & 158.9 & {\it 145.8} \\
bpic2019 & 48.6 & 37.8 & 43.2 & {\it 27.0} \\
\hline
\end{tabular}
}
\label{tab:evaluationB}
\vspace{-2mm}
\end{table}

In Table~\ref{tab:evaluationB},
the effectiveness of the implementation is evaluated in order to understand how the improvements in the implementation contribute to the overall efficiency of the approach.
Columns in the table represent the execution time of the replay approach when no caching, only post-fix caching, only activity caching and the sum of post-fix caching and activity
caching is deployed. In the vast majority of logs, the comparison shows that ITBR provides the best performance.

\vspace{-2mm}
\subsection{Evaluation: Comparison Between Fitness Values}
\label{sec:comparisonFitnessValues}

\begin{table}[ht]
\vspace{-3mm}
\caption{Fitness values comparison between the considered approaches on models extracted by the alpha miner and the inductive miner.
Here, the first column is the name of the log, from the second to the seventh there is the fitness (whether the algorithms succeed) calculated for the different approaches on models extracted by the inductive miner, from the eight to the tenth there is the fitness calculated by the different considered token-based replay approaches on models extracted by the alpha miner.}
\centering
\resizebox{\textwidth}{!}{%
\begin{tabular}{|l|cccccc|ccc|}
\hline
~ & \multicolumn{6}{|>{}c|}{Inductive Miner} & \multicolumn{3}{|>{}c|}{Alpha Miner} \\
\hline
{\bf Log} & {\bf ITBR} & {\bf CTBR} & {\bf ABR} & {\bf BTBR} & {\bf AFA} & {\bf REABR} & {\bf ITBR} & {\bf CTBR} & {\bf BTBR} \\
\hline
repairEx & 1.0 & 1.0 & 1.0 & 1.0 & 1.0 & 1.0 & 0.88 & 0.88 & 0.88 \\
reviewing & 1.0 & 1.0 & 1.0 & 1.0 & 1.0 & 1.0 & 1.0 & 1.0 & 1.0 \\
bpic2017(offer) & 1.0 & ~ & 1.0 & 1.0 & 1.0 & 1.0 & 0.72 & ~ & 0.72 \\
receipt & 1.0 & 1.0 & 1.0 & 1.0 & 1.0 & 1.0 & 0.39 & 0.39 & 0.39 \\
roadtraffic & 1.0 & ~ & 1.0 & ~ & 1.0 & 1.0 & 0.62 & ~ & 0.62 \\
Billing & 1.0 & ~ & 1.0 & 1.0 & 1.0 & 1.0 & 0.69 & ~ & 0.69 \\
\hline
\end{tabular}
}
\label{tab:evaluationC}
\vspace{-2mm}
\end{table}

In Table~\ref{tab:evaluationC}, a comparison between the fitness values recorded by the ITBR, the CTBR and the ABR is provided, for both the alpha miner and inductive miner models. From then onwards, the biggest logs (bpic2017(application), bpic2018, bpic2019) are dropped, since a qualitative evaluation is performed.
For some real-life logs (bpic2017, roadtraffic, Billing) the CTBR did not succeed in the replay in a reasonable time (an empty space has been
reported in the corresponding columns). Alignments have not been evaluated on the models extracted by the alpha miner since it is not assured to have a sound workflow net to start with.
The fitness values obtained in Table \ref{tab:evaluationC} show that the ITBR, on these logs and the models extracted from them by the inductive miner,
is as effective in exploring invisible transitions as the CTBR and the ABR.

\vspace{-2mm}
\subsection{Evaluation: Comparison Between Outputs}
\label{sec:outputsComparison}

\begin{table}[ht]
\vspace{-8mm}
\caption{Comparison between the output of the ITBR and the ABR. First, the name of the log is reported. Then, the number of transitions activated by the two methods is reported, and some aggregations of the similarity measure are provided. Rightmost, the fitness values are reported.}
\centering
\resizebox{0.75\textwidth}{!}{%
\begin{tabular}{|l|cc|ccc|cc|}
\hline
~ & \multicolumn{2}{|>{}c|}{Transitions} & \multicolumn{3}{|>{}c|}{Similarity} & \multicolumn{2}{|>{}c|}{Fitness} \\
\hline
{\bf Log} & {\bf ABR} & {\bf ITBR} & {\bf Min} & {\bf Max} & {\bf Med} & {\bf ABR} & {\bf ITBR} \\
\hline
repairEx & 34858 & 30459 & 0.75 & 1.0 & 0.94 & 0.934 & 0.941 \\
reviewing & 9412 & 8912 & 0.81 & 1.0 & 0.937 & 0.967 & 0.974 \\
bpic2017(offer) & 257970 & 258565 & 1.0 & 1.0 & 1.0 & 0.995 & 0.996 \\
receipt & 27375 & 26642 & 0.42 & 1.0 & 0.94 & 0.839 & 0.863 \\
roadtraffic & 1184482 & 1023901 & 0.35 & 1.0 & 0.625 & 0.791 & 0.816 \\
\hline
\end{tabular}
}
\label{tab:evaluationD}
\vspace{-2mm}
\end{table}

A comparison between the output of token-based replay and alignments has been proposed in Table \ref{tab:evaluationD}.
Some popular logs, that are taken into account also for previous evaluations, are being filtered in order to discover a model (using the inductive miner)
that is not perfectly fit against the original log. Instead of comparing the fitness values, the comparison is done on the similarity between the set of transitions that are activated in the model
during the alignments and the set of transitions that are activated in the model during the token-based replay. The more similar the two sets are, the higher the value of similarity should be.
The similarity is calculated as the ratio of the size of the intersection of the two sets and the size of the union of the two sets. This is a simple approach, with some limitations:
1) transitions are counted once during the replay 2) the order in which transitions are activated is not important 3) the number of transitions activated by the alignments is intrinsically
higher: while token-based replay could just insert missing tokens and proceed, alignments have to find a path in the model from the initial marking to the final marking, so a higher number of transitions is expected.
This comparison, aside fitness values, confirm that the results of the two replay operations, that is a set of transitions activated in the model, are similar.
Table \ref{tab:evaluationD} provides some further evidence that the two replay techniques are comparable.

\vspace{-3mm}
\subsection{Evaluation: Handling of the Token-Flooding Problem}
\label{sec:tokenFloodingSec}

\begin{table}[ht]
\vspace{-8mm}
\caption{Handling of the token-flooding problem: evaluation between outputs with ($\textit{ITBR}^{\textit{+TFC}}$) and without token-flooding cleaning. With the approach enabled, more similar results
to alignments are obtained. In the table, the fitness values are reported. Then, in the middle columns, the number of transitions enabled by the methods are inserted. Eventually, the median of the similarity values, as in Section \ref{sec:outputsComparison}, is reported.}
\resizebox{\textwidth}{!}{%
\begin{tabular}{|l|ccc|ccc|cc|}
\hline
~ & \multicolumn{3}{|>{}c|}{Fitness} & \multicolumn{3}{|>{}c|}{Transitions} & \multicolumn{2}{|>{}c|}{Similarity} \\
\hline
{\bf Log} & {\bf ABR} & {\bf ITBR} & {\bf ITBR$~^{\textrm{+TFC}}$} & {\bf ABR} & {\bf ITBR} & {\bf ITBR$~^{\textrm{+TFC}}$} & {\bf ITBR} & {\bf ITBR$~^{\textrm{+TFC}}$} \\
\hline
repairEx & $0.934$ & $0.941$ & $0.934$ & $34858$ & $30459$ & $30459$ & $0.94$ & $0.94$ \\
reviewing & $0.967$ & $0.974$ & $0.967$ & $9412$ & $8912$ & $8912$ & $0.937$ & $0.937$ \\
bpic2017(offer) & $0.995$ & $0.996$ & $0.995$ & $257970$ & $258565$ & $259597$ & $1.0$ & $1.0$ \\
receipt & $0.839$ & $0.863$ & $0.862$ & $27375$ & $26642$ & $27508$ & $0.94$ & $0.94$ \\
roadtraffic & $0.791$ & $0.816$ & $0.791$ & $1184482$ & $1023901$ & $1184039$ & $0.625$ & $0.625$ \\
\hline
\end{tabular}
}
\label{tab:tokenFlooding}
\vspace{-2mm}
\end{table}

In Table \ref{tab:tokenFlooding}, the importance of handling the token flooding problem is illustrated on several logs.
The models against which the technique is evaluated are the same obtained in section \ref{sec:outputsComparison}.
For both the fitness values (albeit the underlying concepts/fitness formulas are different) and the number of transitions activated in the model,
we are getting a more similar (higher) number, since the activation of unwanted parts of the process model is avoided.
For the median of similarity between the outputs, we obtain equal numbers between the ITBR and the ITBR$~^{\textrm{+TFC}}$ approach; this means that the token flooding
procedure acts only on the most problematic traces of the log according to the model.

\begin{figure*}[!t]
\centering
\begin{lstlisting}[language=Python, basicstyle=\scriptsize, frame=single, numbers=left]
from pm4py.objects.log.importer.xes import factory as xes_importer
from pm4py.algo.discovery.alpha import factory as alpha_miner
from pm4py.algo.conformance.token_replay import factory as tr_factory
log = xes_importer.apply("C:\\running-example.xes")
net, im, fm = alpha_miner.apply(log)
aligned_traces = tr_factory.apply(log, net, im, fm)
\end{lstlisting}
\caption{PM4Py code to load a log, apply the alpha miner and visualize a Petri net.}
\label{fig:examplePm4py}
\end{figure*}

\section{Tool Support}
\label{sec:tool}

The contribution described in this paper has been implemented in the Python library PM4Py.
The tool can be easily installed in the Python 3.7 environment following the documentation reported on the website.
The application of token-based replay is performed
on an event log and an accepting Petri net.
Example code to import a XES file, apply the alpha miner and then the token-based replay is presented in Fig. \ref{fig:examplePm4py}.

In the tool, we provide also some advanced diagnostics, in order to be able to answer to the following questions:
\begin{enumerate}
\item If a given transition is executed in an unfit way, what is the effect on the throughput time?
\item If a given transition is executed in an unfit way, why does this happen?
\item If a given activity that is not contained in the process model is executed, what is the effect on the throughput time?
\item If a given activity that is not contained in the process model is executed, why does this happen?
\end{enumerate}
For questions 1) and 3), the {\it throughput time} diagnostic introduced in Section \ref{sec:diagnostics} can be used.
For questions 2) and 4), the {\it root cause analysis} diagnostic introduced in Section \ref{sec:rootCauseAnalysis} can provide the corresponding answers.

The documentation about the usage of the token-based replay\footnote{\url{http://pm4py.pads.rwth-aachen.de/documentation/conformance-checking/token-based-replayer/}} and of the diagnostics\footnote{\url{http://pm4py.pads.rwth-aachen.de/documentation/conformance-checking/token-based-replayer/token-based-replay-diagnostics/}} is available on the website.

\subsection{Advanced Diagnostics: Throughput Time Analysis}
\label{sec:diagnostics}

The comparison between the throughput time in non-fitting cases and fitting cases permits to understand, 
for each kind of deviations, whether it is important or not important for the throughput time.
To evaluate this, the ``Receipt phase of an environmental permit application process'' log is taken. 
After some filtering operations, the model represented in Figure~\ref{fig:receiptFiltered} is obtained.
Several activities that are in the log are missing according to the model, while some transitions have fitness issues.
After doing the token-based replay enabling the local information retrieval, and applying the {\it duration\_diagnostics.diagnose\_from\_trans\_fitness} function to the log
and the transitions fitness object, it can be seen that transition {\it T06 Determine necessity of stop advice} is executed in an unfit way in $521$ cases.
For the cases where this transition is enabled according to the model the median throughput time is around $20$ minutes, while in the cases where this transition is executed in an unfit way the median throughput time is $1.2$ days.
So, the throughput time of unfit cases is $146$ times higher in median than the throughput time of fit cases.
Activities of the log that are not in the model are likely to make the throughput time of the process higher since they are executed rarely.
In our implementation, applying the \\ {\it duration\_diagnostics.diagnose\_from\_notexisting\_activities} method, the median execution time of cases containing these activities can be retrieved and compared with the median execution time of cases that do not contain them (that is $20$ minutes).
Taking into account the activity {\it T12 Check document X request unlicensed}, it is contained in $44$ cases, which median throughput time is $6.9$ days ($505$ times higher than the standard).

\subsection{Advanced Diagnostics: Root Cause Analysis}
\label{sec:rootCauseAnalysis}

Root cause analysis is a type of diagnostic, that is obtained on top of the token-based replay results, that permits to understand the reasons why a deviation happened.
This is done using the ideas of the framework described in \cite{de2014general}:
\begin{itemize}
\item Log attributes (at the case and the event level) are transformed into
numeric features (for string attributes, one-hot encoding is applied); for each case, a vector of features is obtained.
\item A class (e.g. for (2), 0 for fit traces, 1 for unfit traces; for (4), 0 for traces not containing the activity, 1 for traces containing it)
is assigned to each case.
\item A machine learning algorithm is applied to learn a representation of the data.
\end{itemize}
By transforming the log into a matrix of numeric features, interoperability is kept across a wide set of machine learning classification algorithms (e.g. decision trees, random forests, deep learning methods).
Within our implementation, decision trees are used to get a description of the differences between the two classes.
The decision tree in our approach was trained on the entire dataset, since the goal is to obtain some discrimination rules between the two classes.

\begin{figure}[ht]
\vspace{-4mm}
    \begin{minipage}{0.47\textwidth}
\resizebox{\columnwidth}{!}{%
\includegraphics[width=200px]{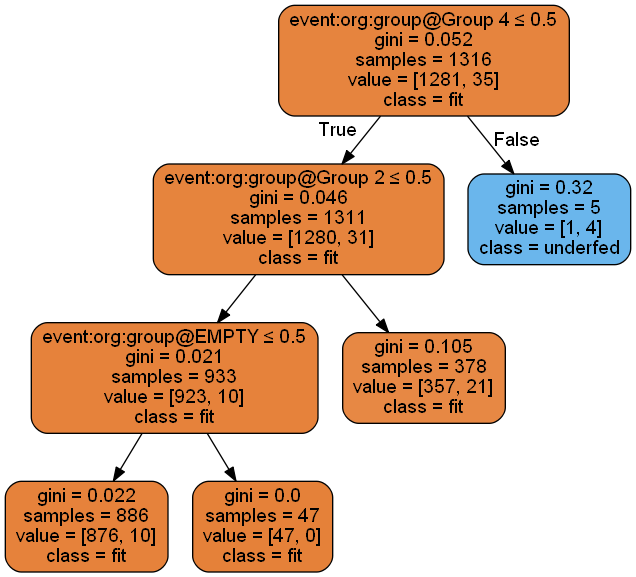}
} \\
{\bf a)} Decision tree extracted comparing fit and unfit cases for transition {\it T02 Check Confirmation of receipt}.
    \end{minipage}
	\begin{minipage}{0.06\textwidth}
	~
	\end{minipage}
    \begin{minipage}{0.47\textwidth}
\resizebox{\columnwidth}{!}{%
\includegraphics[width=200px]{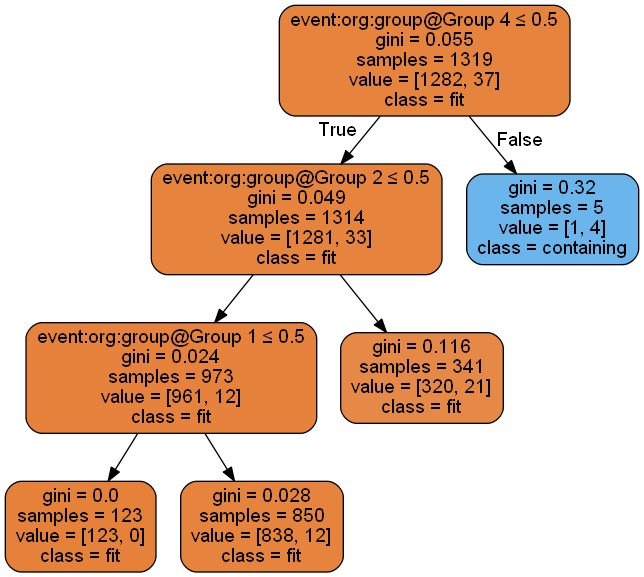}
} \\
{\bf b)} Decision tree extracted comparing cases containing and not containing activity {\it T03 Adjust confirmation of receipt} that is not in the model.
\end{minipage}
\caption{Root Cause Analysis performed on the log ``Receipt phase of an environmental permit application process'' and a model extracted using inductive miner on a filtered version of the log.
In the represented decision trees, two different kind of deviations, a) an activity that is in the model but is executed in an unfit way, and b) an activity is executed that is not in the model, have been analyzed.}
\label{fig:dectreet}
\vspace{-4mm}
\end{figure}

This framework permits to answer to the following questions:
\begin{enumerate}
\item If a given transition is executed in an unfit way, why does this happen?
\item If a given activity that is not contained in the process model is executed, which is the effect on the throughput time?
\end{enumerate}

To evaluate this, the ``Receipt phase of an environmental permit application process'' log and the model represented in Figure~\ref{fig:receiptFiltered} are taken.
In the following examples, the decision tree has been built using only the {\it org:group} attribute.
Applying the {\it root\_cause\_analysis.diagnose\_\allowbreak from\_\allowbreak notexisting\_activities} method,
for transition {\it T02 Check Confirmation of receipt} the decision tree shown in Figure~\ref{fig:dectreet}(a) is obtained, that permits to understand the following information:
(1) Group 4 triggers an unfit execution according to the model. (2) Group 2 triggers sometimes an unfit execution according to the model.
Applying the {\it duration\_diagnostics.diagnose\_from\_notexisting\_activities} method,
for activity {\it T03 Adjust confirmation of receipt} the decision tree shown in Figure~\ref{fig:dectreet}(b) is obtained, that permits to understand
that Group 4 and 2 trigger the activity.

\vspace{-3mm}
\section{Related Work}
\label{sec:relatedWorkSection}

\begin{table}[!t]
\vspace{-2mm}
\centering
\caption{A description of the replay techniques presented in section \ref{sec:relatedWorkSection}. The third column is the target model.
The fourth column describes the super-class of the replay technique (Ali=alignments, TR=token-based replay, DFA=DFA semantics, FP=footprints). The fifth column describes whether the technique is an online technique. The sixth column describes whether the output is optimal (see the bottom of section \ref{sec:relatedWorkSection}).}
\resizebox{\textwidth}{!}{%
\begin{tabular}{|p{1.2cm}|p{5.2cm}|p{1.4cm}|p{1.1cm}|p{1.4cm}|p{1.1cm}|p{1.1cm}|}
\hline
{\bf Refs.} & {\bf Description} & {\bf Model} & {\bf Appr.} & {\bf Online} & {\bf Opt.} \\
\hline
\cite{rozinat2008conformance,de2007genetic,berti2019reviving} & Token-based replay approaches & Petri nets & TR & No & No \\
\hline
\cite{vanden2014event} & Flexible conformance checking approach, based on a decomposition & Petri nets & TR/Ali & Yes & No \\
\hline
\cite{adriansyah2011cost} & Alignments with optimal cost & Petri nets & Ali & No & Yes \\
\hline
\cite{lee2018recomposing} & Replay technique based on decomposing the model, performing alignments and recomposing the result & Petri nets & Ali & No & Yes/No \\
\hline
\cite{taymouri2018evolutionary,bauer2019estimating} & Different replay techniques based on alignments approximation & Petri nets & Ali & No & No \\
\hline
\cite{van2012distributed,shugurov2016applying} & Distributed alignments computation & Petri nets & Ali & No & Yes \\
\hline
\cite{van2019online} & Online conformance checking & Petri nets & Ali & Yes & No \\
\hline
\cite{reissner2017scalable} & Alignments on top of automatons & DFA & Ali & No & Yes \\
\hline
\cite{leemans2018scalable} & Technique to verify the fitness of traces on top of process trees through conversion to a finite automaton & DFA & DFA & No & No \\
\hline
\cite{reissner2019scalable} & Alignment technique that exploits a decomposition of the original Petri net model in S-components, converts them in finite automatons, and apply alignment on the single components. & DFA & Ali & No & No \\
\hline
\cite{molka2014conformance} & Replay technique on top of BPMN & BPMN & TR & No & No \\
\hline
\cite{der2016data} & Footprints comparison & Any & FP & No & No \\
\hline
\end{tabular}
}
\label{tab:relatedWork}
\end{table}

Token-based replay has been introduced as a conformance checking technique in \cite{rozinat2008conformance}.
The approach has also been used internally in some process discovery algorithms such as the genetic miner \cite{de2007genetic} to evaluate the quality of the candidates.
Recently, a flexible online replay technique, that provides token-based replay as option, has been described in \cite{vanden2014event}.
This is based on a decomposition of the model, in such way the state space exploration can be performed with better performance. The approach introduced in this paper has been compared against \cite{rozinat2008conformance}; in comparison to \cite{vanden2014event}, our approach does not require a decomposition of the model.

Another conformance checking technique for Petri nets is the one of {\it footprints} \cite{der2016data}. In this technique, a footprint table is found on both the process model (describing the relationships between the activities as in the model)
and the event log (describing the relationships between the activities as recorded in the process execution). Then, a comparison is done between these two tables.
While this technique is very scalable for conformance checking, it is not a proper replay technique as it does not provide a sequence of transitions in the model.

Currently, the standard replay technique on Petri nets is the computation of {\it alignments with optimal cost} \cite{der2016data,conf-check-book-2018}. In the assessment, we have compared against the approach described in \cite{adriansyah2011cost}, showing that our token-based replay provides better performance than such technique.

Other techniques are based on decomposing the model \cite{van2012decomposing,munoz2014single}, in order to perform a multiple number of smaller alignments.
The recomposition approach described in \cite{lee2018recomposing} is able to provide the optimal cost of an alignment between the model and the process execution under some assumptions. The technique usually leads to a shorter execution times.
However, token-based replay is often still faster (as shown in the assessment).

Approaches to approximate the conformance checking results are described in \cite{taymouri2018evolutionary,bauer2019estimating}; these might not produce the optimal
cost of an alignment but produce generally a good approximation of the alignment or of its cost.
In comparison, our approach is able to produce a proper path in the model when the execution is fit (see the assessment).

In \cite{van2012distributed,shugurov2016applying}, map-reduce approaches have been applied to parallelize the computation of the alignments.
Online conformance checking techniques \cite{van2019online} iteratively update the alignment to include new events; in doing so, for efficiency reasons, the number of states stored and visited might be reduced,
hence optimality of the alignments is not granted. The improved token-based replay approach introduced in the paper is an offline technique. At the moment, we don't provide any scalable map-reduce architecture.

Other replay techniques have focused on different types of process models. In \cite{leemans2018scalable}, a process tree discovered using inductive miner is converted into a deterministic finite automaton
for fast fitness checking. In \cite{reissner2017scalable}, the goal is to perform alignments on automatons. This shows some advantages in models without concurrency, but suffer from scalability issues in models with concurrency. 
In \cite{reissner2019scalable}, a decomposition of a Petri net model into S-components is performed in order to get a collection of automatons, against which alignments are performed.
In \cite{molka2014conformance}, an efficient replay technique for BPMN models is proposed.

Table \ref{tab:relatedWork} summarizes the approaches discussed in this section. The optimality concept is defined only for the techniques producing
alignments (see \cite{conf-check-book-2018}).
Since token-based replay techniques are based on heuristics for invisible/duplicate transitions, and the footprints technique is a
matrix comparison, they have been considered as non-optimal.

\vspace{-3mm}
\section{Conclusion}
\label{sec:concl}
In this paper, an improved token-based replay approach for Petri nets has been proposed.
The technique exploits a preprocessing step that leads to a better handling of invisible transitions. Moreover, the intermediate storage techniques have been improved to achieve a lower execution time.

Token-based replay approaches already outperformed alignment-based approaches for Petri nets with visible transitions.
The proposed token-based replay approach is faster than alignment-based approaches for Petri nets also for models with invisible transitions.

Next to an increase in speed, the problem of token flooding is addressed by ``freezing'' superfluous tokens (see Section \ref{sec:tokenFlood}). 
In this way, the replay does not lead to markings with many more tokens than what would be possible according to the model,
avoiding the activation of unwanted parts of the process models and leading to lower values of fitness for problematic parts of the model.
Moreover, we showed that we are able to diagnose the effects of deviations on the case throughput time, and we are able to perform root cause analysis.

The approach has some limitations. First, we do not propose any termination or fitness guarantees.
Also, performance is in some cases worse 
than advanced replay techniques as automaton-based alignments (as AFA). However, the improved token-based replay has a clear performance lead on the biggest
logs and models that have been considered (BPI Challenge 2017, 2018 and 2019).

We hope that this will trigger a revival of token-based replay, a technique that seemed abandoned in recent years.
Especially when dealing with large logs, complex models, and real-time applications, the flexible tradeoff between quality and speed provided by
our implementation is beneficial.

{\bf Acknowledgements}:
We thank the Alexander von Humboldt (AvH) Stiftung
for supporting our research.

\bibliographystyle{splncs04}
\bibliography{tokenbasedreplay}

\end{document}